\documentclass[a4paper]{article}

\usepackage{amsmath,amssymb,amsfonts}%

\usepackage{multirow}%
\usepackage{array} 

\usepackage[utf8]{inputenc}
\usepackage[T1]{fontenc}
\usepackage{lmodern}
\usepackage{textcomp}

\usepackage{xcolor}
\usepackage{soul}

\usepackage{graphicx}
\usepackage{epstopdf, epsfig}
\usepackage[lofdepth,lotdepth]{subcaption}

\newcommand{\der}[2]{\frac{d#1}{d#2}}
\newcommand{\pder}[2]{\frac{\partial#1}{\partial#2}}

\newcommand{\oforder}[1]{{\cal O}\left(#1\right)}

\newcommand{\air}{}
\newcommand{\wat}{^{(w)}}
\newcommand{\Uwind}{{\mathcal{U}}}

\newcommand{\coeff}[1]{\ensuremath{\mathcal{#1}}}
\newcommand{\Pcoeff}{\coeff{P}}
\newcommand{\Hcoeff}{\coeff{H}}
\newcommand{\Xcoeff}{\coeff{X}}

\newcommand{\Icoeff}{\coeff{I}}
\newcommand{\Io}{\Icoeff_{0}}

\newcommand{\zo}{z_{\emptyset}}  
\newcommand{\zot}{\tilde{z}_{\emptyset}}  
\newcommand{\tho}{{\theta_0}}

\newcommand{\ie}{\textit{i.e.} }

\begin{document}

\title{A generic framework for extending Miles' approach to wind-wave interactions}

\author{
  \parbox{\textwidth}{
    Christophe Chaubet$^1$\footnote{christophe.chaubet@umontpellier.fr} ,
    Miguel M. Manna$^1$\footnote{miguel.manna@umontpellier.fr} ,
    Norbert Kern$^1$\footnote{norbert.kern@umontpellier.fr} 
    \\[0.5em]
    1 - L2C, Université Montpellier, CNRS, Montpellier, France
    }
}

\maketitle

\begin{abstract}
  Understanding the energy transfer from wind to waves is an important but complex topic, typically based on phenomenology, on heavy numerics, or on technically challenging analysis, carried out on a case by case basis. Here we show that the approach by Miles, initially proposed for a still and infinitely deep ocean of inviscid water, is in fact generic: it can easily be adapted, as we demonstrate directly from the mathematical structure of the arguments put forward by Miles. We establish simple transformations, which infer wave growth rates in complex hydrodynamic situations directly from those in Miles' conditions, without any adjustable parameters. The corresponding conversion factors are determined from the hydrodynamic water pressure produced by a propagating surface wave, and  can be determined {\it without} requiring to further analyse wind and air flow.
  Alternatively, the dispersion relation (in the absence of air andt hus wind) may be used as a starting point. We reproduce a variety of results for different hydrodynamic situations to show how such generalisations can be achieved with surprisingly simple calculations and without any additional numerical effort, which should make the approach interesting for real-life applications.
\end{abstract}


\section{Introduction}

How do ocean waves grow under the effect of wind? This perfectly straightforward question turns out to be remarkably difficult to answer in detail. Several approaches have been proposed historically \cite{Jeffreys:1924,Jeffreys:1926,Munk:1947,Phillips:1957,Miles:1957}, but the one proposed by Miles \cite{Miles:1957} is particularly appealing in that it can be formulated entirely based on fluid mechanics, without invoking {\it ad hoc} arguments. Today, it is one of the pillars for understanding ocean waves, and continues to receive attention \cite{Young:2014,Carpenter:2017,Bonfils:2022}. 
\\

In practice, however, the approach poses formidable challenges in terms of the calculations it requires, especially since it is usually posed as a problem involving two distinct hydrodynamic domains, the water and the air, separated by a free interface on which the wave propagates.The issue is therefore to coherently solve the hydrodynamics in the air domain and in the water domain, coupled through the appropriate boundary conditions at the (evolving) interface. This
task remains formidable even after linearisation. It has been completed by Miles \cite{Miles:1957}, assuming the simplest possible scenario, in which the wave propagates on an infinitely deep and otherwise static ocean, without any currents. Both water and air are considered inviscid and incompressible.  Miles has proposed a formulation of this problem which leads to the Rayleigh equation \cite{Rayleigh:1880}, and Beji \& Nadaoka \cite{Beji:2004} have put forward a numerical procedure to solve the integrals required to deduce the wave growth coefficients. Generalisations to more complex hydrodynamic scenarios have only been introduced and revisited more recently, accounting for finite water depth \cite{Montalvo:2013}, shear flow mimicking underwater currents \cite{Kern:2021}, and the role of viscosity \cite{Miles:1959:shearII,Chaubet:2024}. All of these modifications are of direct interest for practical applications, such as the forecast of ocean waves \cite{Young:book,Bretschneider:1951,Bretschneider:1952,Neumann:1952,Neumann:1954,PiersonJr:1955}, the growth of waves in the shore regions \cite{Young:book,Young:1996,Young:1996-II,Donelan:2005,Donelan:2006,Young:2005}
and the associated issues of coastal erosion \cite{Dean:book}, but potentially also for questions involving other issues such as the installation of sea-based wind farms \cite{Allaerts:2019,Porchetta:2021,Akhtar:2021,Ma:2022,Maas:2023}.

Miles’ approach is particularly appealing in that it gives access to the physical mechanisms. It is, however, intimidating, as it cumulates the technical difficulties of matching the flow between two sub-systems, a free interface, boundary layer physics, etc. Indeed, each of the studies generalising the approach to other hydrodynamic situations has followed the path set out by Miles, solving the two fluid problem for a propagating wave. But these calculations may in all cases be described as lengthy, and probably sufficiently so to discourage from further generalisations in view of exploring other, yet more complex hydrodynamic situations.

Here we show that the approach is in fact simpler to generalise than one might expect. Building on the work by Miles, we set up a completely generic mapping which extends Miles' results on the impact of wind to more complex, more realistic hydrodynamic situations. It is based exclusively on analysing a surface wave {\it without} considering air or wind, and establishes coefficients which rigorously characterise the underlying hydrodynamics. By simple multiplication to Miles' results this then establishes the growth coefficients in the complex system. This procedure is simple and straightforward,  without need for any additional numerical work, yet rigorous and totally generic.
\\

\begin{figure}
   \center
  \begin{subfigure}{0.45\textwidth}
    \center
    \includegraphics[width=2.0\textwidth]{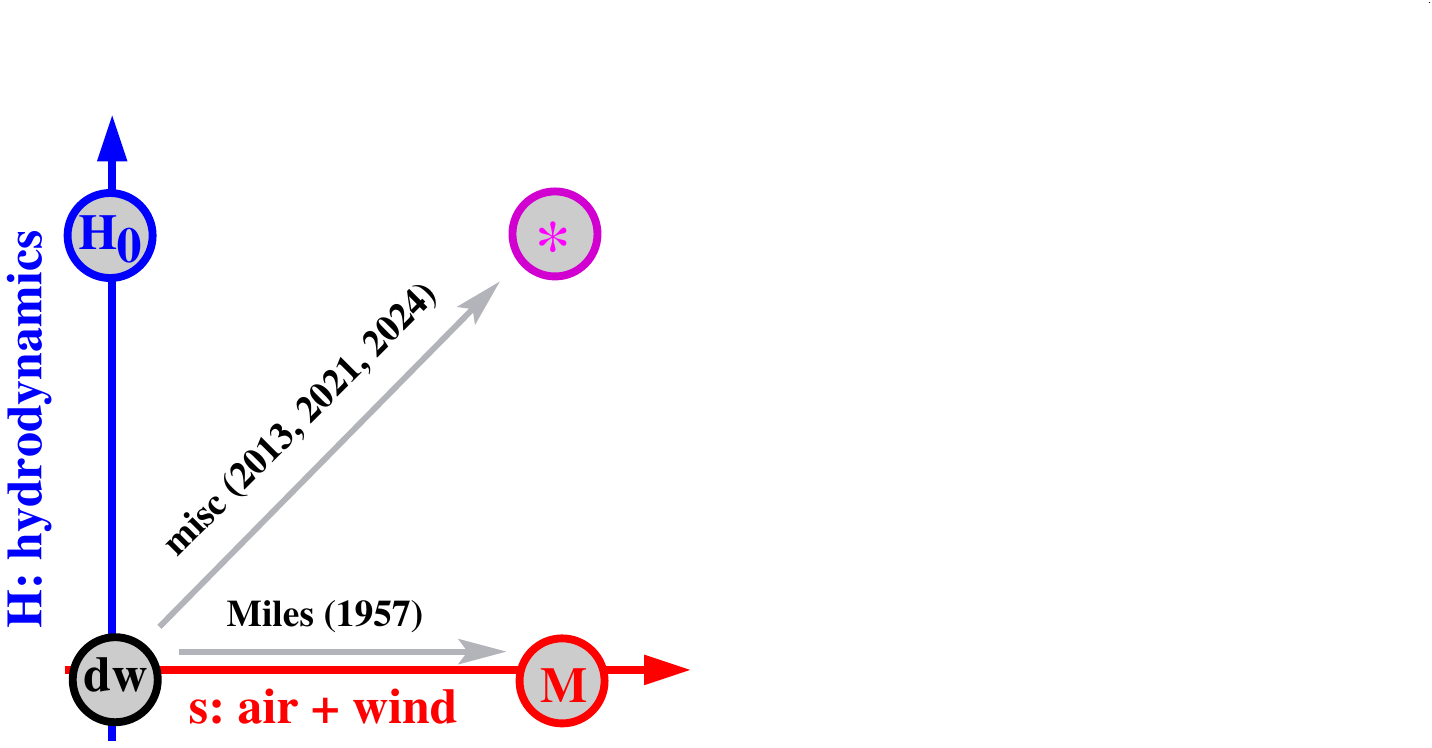}
    \subcaption[]{}
  \end{subfigure}
  \hfill
  \begin{subfigure}{0.45\textwidth}
    \center
    \includegraphics[width=2.0\textwidth]{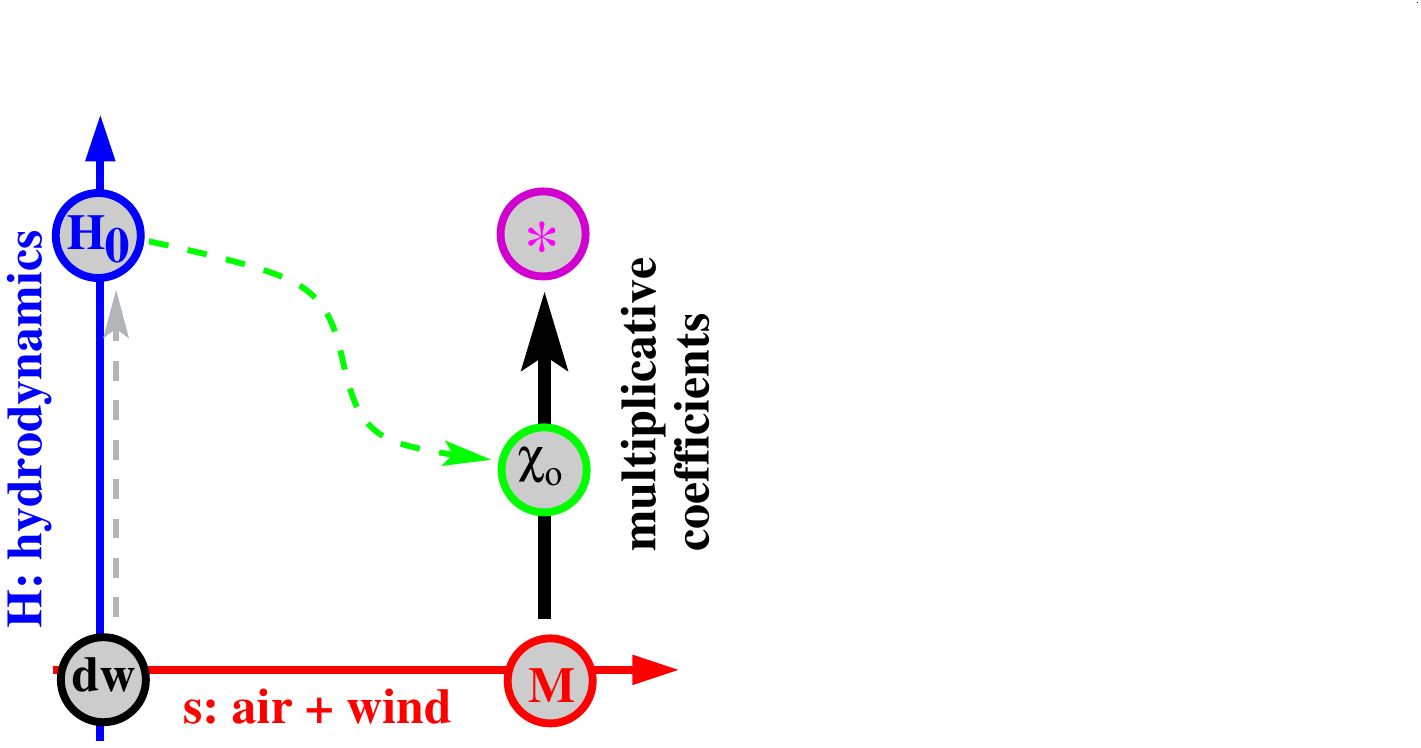}
    \subcaption[]{}
  \end{subfigure}
  \caption{\label{fig:sketch}
    (a) Miles' work starts from gravity waves in deep water ('dw') and adds the effect of air motion including wind. Various further studies \cite{Montalvo:2013,Kern:2021,Chaubet:2024} have followed a similar approach to generalise the analysis to different hydrodynamic conditions, symbolised as 'H' (finite water depth, currents, viscosity, etc.). Calculations are tedious.
    (b) The framework proposed here proceeds differently. First, the hydrodynamic analysis is required for gravity waves in the generalised hydrodynamic conditions, for a system (H$_0$) {\it without} air and wind. From this follow multiplicative coefficients such as $\Xcoeff_0$ which, when applied to Miles' initial results ('M'), yield growth coefficients due to wind, for the generalised hydrodynamic conditions ('H'). Calculating and applying the coefficients is straightforward, no further numerical work is required.
  }
\end{figure}

Our approach is summarised schematically in the sketch shown in Fig.~\ref{fig:sketch}, which will also help to visualise the structure of the article. 
Section \ref{sec:outline} summarises Miles' route to wave growth under the effect of wind for a deep and still ocean, while also introducing the notations we use.
Section \ref{sec:framework} presents our approach to formalise the hydrodynamic conditions in a very generic way. It then shows how describing the wave propagation in these conditions - but {\it without} accounting for air and wind yet - is ultimately sufficient to quantify the impact of wind, following the spirit of Miles' analysis. Specifically, multiplicative coefficients are established from the pressure perturbations.
Section \ref{sec:examples} validates this approach by reproducing, quite effortlessly, previous results for various hydrodynamic situations.
Section \ref{sec:others} extends the results to the effect of wind on the propagation speed as well as Miles' pressure coefficients, and shows another shortcut, which requires only the dispersion relation (in the airless, windless system) as a starting point.
Section \ref{sec:discussion} summarises our findings and discusses how the framework we put forward here should prove useful.

\section{Outline of Miles' approach} \label{sec:outline}

Here we outline Miles' approach, summarising its spirit and sketching the arguments. Tthis section contains no new results, but aims to provide a recap of Miles' historical approach, in a way which prepares our arguments for going beyond those results. However, we present the analysis in a way which exposes its structure, independent of any specific hydrodynamic scenario: this is what will make a generic approach possible. As full mathematical derivations are available in the literature \cite{Miles:1957,Beji:2004}, we focus on the essence of the arguments here, rather than reproducing all derivations in detail.

\subsection{The essence of the approach by Miles}
\label{sec:essence}

In the spirit of the model by Miles we are thus dealing with two fluids, air and water, coupled via an interface at the vertical position $z=0$. In absence of a wave, the flow fields  $\vec U_{ext}(z)$ and $\vec V_{ext}(z)$ are prescribed in the air and in the water, respectively, translationally invariant both in the $x$ and $y$ directions.
In particular, writing the stationary air flow,  in the positive $x$ direction,
as
\begin{equation}
\label{eq:windprofile}
\vec{U}_{ext}(z) = \Uwind \, f(z) \, \vec e_x    
\end{equation}
introduces the \textit{wind strength} $\Uwind$ and the (external, imposed) \textit{wind profile} $f(z)$.

\paragraph{Notations}
A remark on notations is in order. Different conventions have been used by different authors mentioned above, and sometimes for different manuscripts by the same research groups. Conflicts are therefore unavoidable, and we use this manuscript as a way to put forward a consistent choice of notations, as close as possible to existing ones, but making adjustments when necessary to leave room for the extensions we implement.

For example, in this manuscript we systematically reserve the indices '0' and '1' to the successive terms of a perturbation theory.
Consequently, we use $\Uwind$ for the  {\it wind strength} (rather than the common $U_1$). Also, we refer to the pressure in the absence of a wave as $P_{ext}$ (rather than $P_0$). For the same reason, we use the slightly unconventional notation $\zo$ (rather than $z_0$) for the {\it roughness length} to be introduced in the following paragraphs.

\paragraph{Normal mode analysis}\label{sec:normalmodes}

\begin{figure}
\centering
  \includegraphics[width=0.7\textwidth]{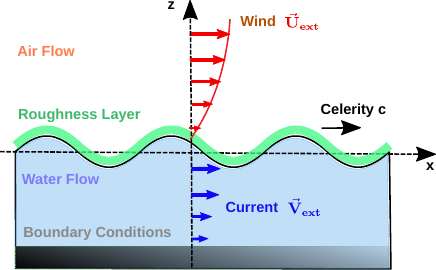}
  \caption{
    Illustration of Miles' framework: in the air domain a profile $U_{ext} = \mathcal{U} \, f(z)$ is imposed to represent the vertical modulation of the (horizontal) wind, aligned with the propagation of the sinusoidal perturbation, with wave number $k$ and amplitude $\eta$ and progressing with celerity $c$. A flow $\vec{V}_{ext}$ may be imposed in the water domain, respecting the boundary conditions at the bottom. Both flow fields are matched at the interface, where one also accounts for the presence of a {\it roughness layer}, of thickness $\zo$, taken to capture all boundary layer effects.
  }
  \label{fig:illustration}
\end{figure}

Surface waves are analysed in terms of normal modes for the dynamic perturbation. As a wave passes, in the $x$ direction, the water-air interface is perturbed to (see Fig.~\ref{fig:illustration})
\begin{equation}
\label{eq:wave}
    z=\eta(x,t) = \eta \, e^{i k (x - c \, t)} = \eta \, e^{i\varphi} \ ,
\end{equation} 
with wave number $k$ and celerity $c$, and $\varphi:=k(x-c \,t)$ being shorthand for the phase. The flow fields are taken to be  $\vec U = \vec U_{ext}(z) + \vec u(x,z,t)$ in the air (density $\rho\air$) and  $\vec V(x,z) = \vec V_{ext}(z) + \vec v(x,z,t)$ in the water (density $\rho\wat$),
where small letters represent the dynamic perturbations due to the wave.
Each flow field obeys the inviscid Navier-Stokes equations with the corresponding densities and appropriate boundary conditions.
In the context initially treated by Miles, the wind profile is logarithmic, the water is infinitely deep, and there is no pre-existing water flow, \ie $\vec V_{ext} = 0$.

The dispersion relation of the system, unknown for now, links the celerity $c$ for a given wave number $k$, and $c$ may be complex. The amplitude growth coefficient can thus simply be obtained from the imaginary contribution to the celerity, as
\begin{equation}
\eta(x,t) = \eta \, e^{i k(x-\Re[c] t)} \, e^{k \, \Im[c] \, t} =: \eta \, e^{i (kx-\omega t)} \, e^{\gamma \, t}
\end{equation}
with
\begin{equation}
  \label{eq:gamma}
  \omega := k \, \Re[c]
  \qquad\mbox{and}\qquad\gamma := k \, \Im[c]
  \ ,
\end{equation}
where $\omega$ is the angular frequency and $\gamma$ is the amplitude growth rate.
  
\paragraph{Wind profile}

The appropriate wind profile described by the function $f(z)$ is subject to debate, and indeed several profiles have been considered by various authors for open seas and for laboratory experiments \cite{Monin:1954,Riley:1982,Komen:1994:inbook,Morland:1993,Beji:2004,Bruch:2021,Villefer:2021}. The approach we present does not rely on any particular wind profile: switching to a different profile can be accomodated by modifying the value of one key quantity in the results \cite{Beji:2004}, but the approach preserves its validity.

However, to remain concrete, we shall consider the specific example of a logarithmic wind profile $U\air(z)$, also used by Miles \cite{Miles:1957}:
\begin{equation}
  \label{Udefinition:logprofile}
  U\air(z) = \Uwind \ln(\frac{z}{\zo})
  \ ,\quad\mbox{with} \ \ 
  \Uwind  = \frac{u_{*}}{\kappa}
  \quad \mbox{and}  \ \ 
  \kappa \approx 0.41
  \ . 
\end{equation}
The phenomenological constants are   the {\it friction velocity} $u_{*}$, the {\it Von K\'{a}rm\'{a}n} constant $\kappa$, and the {\it aerodynamic sea surface roughness}  $\zo$, typically of the order of a millimetre.
This logarithmic profile can be defended on the basis of arguments based on the Reynolds stress tensor and on Prandtl's mixing length approach \cite{Prandtl:1925,Prandtl:1932,Prandtl:1949}.  In the marine context, this approach has been justified by scaling arguments and by solution matching between the near-surface air layer and the geostrophic air layer \cite{Tennekes:1973}, close to the marine boundary layer (see \cite{Garratt:book,Garratt:1994,Garratt:1996}). It has been submitted recently to experimental tests in laboratory conditions \cite{Branger:2022,Villefer:2021}, as well as measurements over the Danish North Sea \cite{Pena:2008}. In practice, it is commonly used to describe the vertical variation of the horizontal mean wind speed within the lowest layer of air.

\paragraph{Roughness layer and roughness length}

An important point which emerges from the analysis of the mechanisms establishing this vertical density profile is that the length scale $z_0$ in the wind profile may also been interpreted as a {\it roughness length}, \ie the vertical extension of a layer in which turbulence regulates the vertical transfer of momentum in the air
This roughness length is furthermore found to obey the {\it Charnock relation} 
 \cite{charnock:1955}
\begin{equation}
\zo =  \alpha_{CH} \,  \frac{u_*^2}{g} \ , 
\end{equation}
where $\alpha_{CH}$ is the {\it Charnock constant}.
Although there is variation in the values found by different authors for $\zo$ \cite{Pena:2008, Branger:2022, Hersbach:2011}), its order of magnitude is always found to be compatible with the value  $\alpha_{CH} \approx 0.0178$, which has also been used by Miles \cite{Miles:1957}.

According to Charnock \cite{charnock:1955}, $\zo$ is the height at which hydrodynamic matching conditions are to be applied. Following Miles \cite{Miles:1957}, we shall exploit this in the following.

\subsection{Miles' description of the air domain}

As we highlight now, one of the keys to Miles' results is to provide a self-contained description of what happens in the air domain.
The air density is $\rho$ and the air flow field is written
\begin{equation}
  \vec U = \vec U_{ext} + \vec u \, e^{i\varphi}
  \ ,
\end{equation}
where $U_{ext}$ is the imposed air flow field, \ie the wind, $\vec u = (u_x,0,u_z)$ is the dynamic perturbation due to the wave and $\varphi=k(x-ct)$ is shorthand for the phase of the propagating wave.

\paragraph{Hydrodynamics in the air domain}

The air pressure is decomposed as 
\begin{equation}
  \label{eq:pressure:decomposition}
  P = P_{ext} - \rho \, g \, z+ p
  \ ,
\end{equation}
where
\begin{equation}
  p=p(z) \, e^{i \varphi}
\end{equation}
is again the dynamic contribution accompanying the wave.

All air flow is assumed to be incompressible and  inviscid, and the pressure perturbation due to the wave must vanish high up in the atmosphere. From the arguments put forward by Miles {\it et al.} \cite{Miles:1957,Conte:1959} we establish that the dynamic air pressure perturbation due to the passing wave is given, for $z>\zo$, as 
\begin{equation}
  \label{eq:pressure:air:Miles}
  p(z)
  \approx k \rho\air \, \eta c^2 \,
  \int_{z}^\infty \left(\frac{U_{ext}(z')}{c} - 1 \right) \, \frac{u_z(z')}{u_z(\zo)} \, k  dz'
  \ .
\end{equation}
The demonstration is summarised in a compact and accessible way in Beji \& Nadaoka \cite{Beji:2004}.

\paragraph{Miles' approach to the air pressure at the interface}
Specifically, to find the air pressure at the interface, Eq.~(\ref{eq:pressure:air:Miles}) must be evaluated at the interface position modified by the roughness height, \ie at $z=\zo+\eta(x,t)$. This dynamic pressure perturbation is to be added to the standard pressure at the interface, $P_{ext}-\rho g \eta \, e^{i \varphi}$.
This yields the complete expression for the air pressure at the interface, at $z=\eta+\zo$:
\begin{equation}
  \label{eq:pressure:air:full}
  \left.P(x,z,t)\right|_{z=\eta}
  = P_{ext} - \rho\air\, g \eta\, e^{i\varphi} +   k \rho\air \, \eta c^2 \, \Icoeff \, e^{i\varphi}
  \ ,
\end{equation}
where we have defined $\Icoeff$, a key quantity, as
\begin{equation}
  \label{eq:Ic:def}
  \Icoeff 
  :=
  \int_{k\zo}^\infty \left(\frac{U_{ext}(kz)}{c} - 1 \right) \, \frac{u_z(kz)}{u_z(k\zo)} \, d(kz)
  \ .
\end{equation}
Note that this expression implies a first-order approximation in $\eta$, whereby the lower bound of the integral may be taken to be $\zo+\eta \approx \zo$.
As an (important) aside, we point out that including the denominator, $u_z(k\zo)$, in this definition  is essential: the velocity profile will be deduced from a linear ordinary differential equation (the Rayleigh equation (\ref{eq:Rayleigh}), to be introduced below). This definition therefore ensures that the actual amplitude of $u_z(kz)$ is of no importance for evaluating this integral.
From Eq.~(\ref{eq:Ic:def}) it is furthermore clear that the coefficient ${\mathcal I}$ is independent of $k$ (since $\zo$ in the lower bound of $\Io$ obeys the Charnock equation Eq.~(\ref{eq:charnock})).

In essence we may summarise these results by stating that the air pressure perturbation at the interface, due to the air flow accompanying the passing wave, may be expressed as
\begin{equation}
  \label{eq:pressure:air:interface}
  \left.p\right|_{z=\eta} = k\,\eta \, \rho c^2 \Icoeff
  \ ,
\end{equation}
where the integral to determine the coefficient $\Icoeff$ can be evaluated numerically as follows.

\paragraph{Rayleigh  equation for the air flow profile\label{par:Orr-Sommerfeld}}

In a first step,  Miles shows that the (linearised) dynamic air flow perturbation $u_z(z)$ obeys the following equation in the air domain:
\begin{equation}
  \label{eq:Rayleigh}
  u_z''(z) - k^2 \, u_z(z) = \frac{ U_{ext}''(z)}{U_{ext}(z) - c}  \, u_z(z) \; ,
\end{equation}
where the primes stand for the derivative, and $u_z(\infty)=0$.
When $c$ is real (which implies a singularity where the denominator vanishes), this is referred to as the  {\it Rayleigh equation}, which we shall primarily be concerned with. However, we return later to the general case where the celerity $c$ has an imaginary part.

In any case, solving this equation for the air velocity $u_z(z)$ is required for evaluating the integral $\Icoeff$.

\paragraph{Numerical approach to solving the air flow: Beji \& Nadaoka}

Solving the Rayleigh equation is a non-trivial task, due to the singularity where $U(z)=c$. Beji and Nadaoka \cite{Beji:2004} have provided a strategy to do this, by transforming the Rayleigh equation to a Riccati equation, then finding an analytical expansion in the vicinity of the singularity. This is then used to initialise a numerical procedure to determine the air velocity profile $u_z(z)$.
From this, for a given wind profile and for a specific celerity $c$, the integral $\Icoeff$ can then readily be determined.

Most importantly, this means that the resulting relation for the growth coefficient as a function of the non-dimensional parameter $c/\Uwind$ is readily available, in numerical form, for the specific, simplified hydrodynamic conditions studied by Miles. In fact, we shall show that this result is in fact sufficient to also determine the impact of wind on wave growth for general hydrodynamic situations with only very little extra effort.

\subsection{Pressure condition at the interface}

Having established an expression for the air pressure $P\air$, and before addressing the question of the water pressure $P\wat$, we first lay out how the dynamics in these separate domains must ultimately couple at the interface. Locally, the air pressure must match the water pressure, up to  (potentially) an interfacial pressure jump $\Delta p$, as it would for example arise from surface tension. It is useful to separate out the pressure perturbation due to the travelling wave,
$P\wat = P_{ext}\wat - \rho\wat g z + p\wat$, as in Eq.~(\ref{eq:pressure:decomposition}), which leads to the pressure balance at the interface
\begin{equation}
  \label{eq:pressurebalance:full}
  P_{ext}\air
  - \rho\air g \eta + \left.p\air\right|_{z=\eta}
  + \Delta p
  =
  P_{ext}\wat
  - \rho\wat g \eta + \left.p\wat\right|_{z=\eta}
  \ .
\end{equation}
Note that, since $p\air$ and $p\wat$ are a response to the wave, and thus proportional to the amplitude $\eta$ of the perturbing wave, the zero-order pressure balance (at $\eta=0$) implies that the constant pressure terms must match ($P_{ext}\air = P_{ext}\wat$).

A further simplification arises from the fact that, in the linear regime, it is sufficient to evaluate the pressure conditions at $z{=}0$ (rather than at $\eta$, the resulting difference being second order). This establishes a convenient expression for the air pressure perturbation at the interface in terms of $\Icoeff$.
Indeed, using Eqs.~(\ref{eq:pressure:air:full}) and (\ref{eq:pressurebalance:full}), the linearised pressure condition, evaluated at $z=0$, reads
\begin{equation}
  \label{eq:pressurebalance}
  - \rho\air g \eta \, e^{i\varphi} +   k \rho\air \, \eta c^2 \, \Icoeff \, e^{i\varphi} + \Delta p \, e^{i \varphi}
  =
  - \rho\wat g \eta  \, e^{i\varphi} + \left.p\wat\right|_{z=0}   \, e^{i\varphi}
  \ ,
\end{equation}
which in a sense captures all the physics of the wave moving at the interface.

\subsection{Hydrodynamics in the water domain}

In order to exploit the pressure balance, Eq.~(\ref{eq:pressurebalance}), we now require an  expression for the pressure perturbation in the water domain which  accompanies the propagating wave. This is specific to a given hydrodynamic context, and this is of course where the hard work lies.
In Miles' seminal paper, for example, the water domain has been considered as static (in the absence of a propagating wave), non-turbulent,  inviscid and of infinite depth.  In more general conditions the flow in the water domain will be more complex, according to the flow properties, the boundary conditions being applied at the bottom of the water domain, and to external flow which may already be present before the wave passes, for example.

But our aim here is to detach ourselves from any particular hydrodynamic context. For now we shall simply assume that the water pressure field is known as
\begin{equation}
  \label{eq:pressure:water}
  P\wat = P\wat_{ext} - \rho\wat g \, z + p\wat(z)\, e^{i k (x-c t)}
  \ ,
\end{equation}
where $p\wat$ is the dynamic pressure perturbation due to a wave mode of small amplitude, propagating with a (for now unknown) celerity $c$ on the initially flat interface.

Solving these water hydrodynamics is of course a challenging task, in particular in the presence of air flow, and possibly wind. However, we show in the following that for generalising Miles' approach it is sufficient to establish the pressure perturbation for a vanishing density in the air domain, \ie in a  simplified system in which the air is replaced by a homogeneous, constant air pressure (and thus in the absence of wind). This relies on the fact that, to leading order, the presence of air and wind can be fully accounted for through the effect they have on the celerity, which in turn yields the growth rate. The following sections lay out this argument in a self-consistent way, without referring to any particular hydrodynamic context in the water domain, which thus makes the approach extremely versatile.

\subsection{Dispersion relation}

In fact, the previous pressure balance already encodes the dispersion relation. To see this it suffices to equate all terms in $e^{i\varphi}$ in Eq.~(\ref{eq:pressurebalance}) as
\begin{equation}
  \label{eq:dispersion}
  (1-s) \, g \eta + k \eta c^2 \, s \Icoeff = \frac{1}{\rho\wat} \, \left(\left.p\wat\right|_{z=0}
  - \Delta p\right)
  \ ,
\end{equation}
which is an implicit equation for $c(k)$. Here, the parameter $s$, defined as
\begin{equation}
  \label{eq:s:def}
  s := \frac{\rho\air}{\rho\wat}
\end{equation}
quantifies the ratio of the densities in air and water, respectively, and plays a major role. For a given value of $k$, this is therefore an implicit equation for the celerity $c$.

\subsection{Resolution strategy}

Solving the dispersion relation Eq.~(\ref{eq:dispersion}) for an expression $c=c(k)$ is typically impossible in full generality. Miles' answer to this problem is a power series expansion in $s$, since  $s \approx 10^{-3}$. The celerity can therefore be expanded as
\begin{equation}
  \label{eq:c:expansion}
  c = c_0 + s \, c_1 + \oforder{s^2}
  \ ,
\end{equation}
and we shall adopt the same notation for all other variables depending on $s$.

Note first that the effect of wind enters as $s \, \Icoeff$ in the above dispersion equation, and it is therefore sufficient, to leading order in $s$, to consider the dispersion relation with $s \, \Icoeff \approx s \, \Icoeff_0$. Here  $\Io$ is thus understood to be the value of the integral $\Icoeff$ evaluated using $c \approx c_0$.
Evaluating the dispersion relation Eq.~(\ref{eq:dispersion}) to leading order ($s=0$) yields the celerity $c_0$ of the wave in absence of air. 
Knowing $c_0$ then also determines $\Io$ (which can be evaluated via analytical approximations, as done by Miles, or numerically, following Beji \& Nadaoka).

But fixing $c_0$ does more, as it effectively also fixes the pressure term $\left.p\wat\right|_{z=0}$ at the interface: although this pressure depends on the presence of air (and wind), for $s\neq0$, it does so only indirectly, via $c_0$. Essentially, this means that the water pressure perturbation at the interface is set by the fact that a wave propagates at a given celerity $c$, irrespectively of how this wave comes about. This is true even when $c \neq c_0$, and thus leads to a condition which self-consistently fixes the correction to the celerity  $c$ when $s$ is small but non-zero. 

These are the main elements of Miles' resolution strategy, and we now recapitulate his argument in the specific case he has considered initially.

\subsection{Example: Miles' gravity waves}

All arguments so far have been generic, as no particular form of the hydrodynamics has been assumed. For illustration we now  consider the simplest possible gravity waves (infinite depth, no currents, small amplitude, no viscosity, no surface tension), which is the case originally analysed by Miles \cite{Miles:1957}.  The corresponding expression for the amplitude of the dynamic pressure is known to be given by \cite{Janssen:book,Kinsman:book} 
$p\wat(z) \approx \rho\wat \, k \eta \, c^2  \, e^{k z}$,
to first order in $\eta$.
Therefore, to linear order,
\begin{equation}
  \label{eq:miles:pressure}
  \left.p_{M}\wat\right|_{z=0}
  = \rho\wat \, k \eta \, c^2
  \qquad\mbox{(Miles' reference conditions)}
  \, .
\end{equation}
For a wave propagating in Miles' conditions, this is to be evaluated at $c=c_M$, to be determined just below.
Injecting Eq.~(\ref{eq:miles:pressure}) into (\ref{eq:dispersion}), taking $\Delta p=0$ as we neglect surface tension effects, we obtain
\begin{equation}
  (1-s) \, \frac{g}{k} = c_M^2 \, \left(1-s \, \Io\right)
  \qquad\mbox{(Miles reference conditions)}
  \, ,
\end{equation}
which for $s=0$ reproduces the well-known dispersion relation for gravity waves as considered by Miles,
\begin{equation}
  \label{eq:c:Miles}
  c_{M,0} := \sqrt{g/k} \qquad \mbox{(Miles reference conditions)}
  \, .
\end{equation}
Substituting $c_M=c_{M,0}+ s\, c_{M,1}$ into Eq.~(\ref{eq:pressurebalance}) and comparing linear terms in $s$ furthermore gives
\begin{equation}
  \label{eq:cM1}
  c_{M,1} = \frac{c_{M,0}}{2} \, \left(\Io - 1 \right)
  \qquad\mbox{(Miles reference conditions)}
  \, ,
\end{equation}
from which we recover
\begin{equation}
c_M = c_{M,0} \, \left(1 + s \, \frac{\Io-1}{2}\right)
\qquad\mbox{(Miles reference conditions)}
\ .
\end{equation}
The expression for $\gamma_M = k \, \Im[c]$ (see Eq.~(\ref{eq:gamma})) in Miles' free ocean conditions
can thus be written as
\begin{equation}
  \label{eq:gamma:Miles}
  \gamma_M
  = \omega_{M,0} \,  \Im\left[s \, \frac{\Io}{2}\right]
  \qquad\mbox{(Miles reference conditions)}
  \ ,
\end{equation}
using that $c_{M,0}$ is real. This reproduces Eq.~(3.10) in Beji \& Nadaoka \cite{Beji:2004}.

\section{Beyond Miles: a framework for quantifying the impact of water hydrodynamics} \label{sec:framework}

Having reproduce Miles' original  results, we now put forward a generic method to calculate growth rates, almost effortlessly, for other models of the hydrodynamic flow in the water domain.
Before, however, several remarks are worth making here, which are central to the success of Miles' approach.

\subsection{Making the argument fully generic}

First, we underline that it has proven sufficient to determine the integral $\Io$, evaluating Eq.~(\ref{eq:Ic:def}) with the celerity $c_0$ obtained in the absence of air and wind ($s=0$). This is a crucial simplification: it is key to making the problem tractable.

Second, $c_0$ can thus be obtained from a simplified 'airless' ($s=0$) system, in which the air domain has been replaced by a simple constant pressure field. This simplification is huge when it comes to solving the corresponding hydrodynamic problem. But also, often such an analysis in an airless (and hence windless) situation is already available in the literature, and can therefore be exploited directly.

Third note that, consequently, $\Io$ is by construction a quantity defined exclusively in the air domain. It does not, therefore, depend on a specific choice for the hydrodynamic conditions in the water domain, beyond the fact that these will affect $c_0$.

Fourth, as already pointed out, the water pressure perturbation at the interface depends on $s$ only through $c$ (which is a function of $s$), but not explicitly. This can indeed be rationalised: the water displacements required to sustain the travelling wave are not affected by the air density. Again, this will prove important.

Essentially, these arguments make the point that the air domain and the water domain can be analysed separately, assuming a certain celerity $c$, which is then fixed self-consistently by the coupling via a pressure balance. All these arguments are totally generic, and in no way specific to Miles' particular hydrodynamic reference conditions.

\subsection{Choice of non-dimensional variables}

We first address the technical issue of choosing non-dimensional variables, which was not central in Miles' analysis but is key to generalising the approach.

As a length scale it is natural to refer to the wave length, which we do by choosing $1/k$ as a reference scale. For the time scale, we make the choice to refer to the period of a wave propagating in the system of interest but {\it without}   air (and hence wind) being present. Other choices are possible, of course, but typically obscure the arguments we intend to expose. Our choice amounts to choosing $1/\omega_0 = 1/(k \, \Re[c_0])$ as a time scale. Implicitly, the velocity scale is $\Re[c_0]$. We thus define the non-dimensional variables, identified by a tilde, as
\begin{equation}
  \tilde{z} = k \, z \ ,
  \qquad
  \tilde{t} =  \omega_0 \, t = k \, \Re[c_0] \, t \ ,
\end{equation}
etc.
For example, $\tilde{\gamma}=\gamma/\omega_0=\gamma/(k \, \Re[c_0])$ is the non-dimensionalised amplitude growth rate.

Thus two non-dimensional quantities are already present, and central to the arguments,
\begin{equation}
  s = \frac{\rho\air}{\rho\wat} \ll 1
  \ , \qquad \mbox{and} \qquad
  \tho := \frac{\Re[c_0]}{\Uwind}
  \ .
\end{equation}
Here $\Re[c_0]$ is the propagation speed of a wave in the system without air or wind, and we shall refer to $\tho$ as the (monochromatic) wave age, and we we provide some important specifics below.

In addition, there are other hydrodynamic parameters, to describe all aspects of the flow in the water domain. We make the {\it choice} to use variables which do {\it not} depend on the wind strength (other choices can be made, and have been made \cite{Montalvo:2013,Chaubet:2024}, but these have proven to be less convenient for our purposes).
We would thus choose, for example, $\tilde{h} = k  h$ for a finite water depth $h$, $\tilde\Omega = \Omega/(k \, \Re[c_0])=\Omega/\omega_0$ for a bulk shear rate $\Omega$, and $\tilde{\nu} = k \, \nu / \Re[c_0] = k^2 \, \nu / \omega_0$ for a bulk viscosity $\nu$. An important point is that some of these non-dimensional parameters may depend on $\Re[c_0]$, which will need to be accounted for in the following sections.

\subsection{The monochromatic wage age}

The quantity $\theta = c/\Uwind$ has arisen naturally, which is a consequence of choosing the scales of the wave (rather than the scalesof the wind profile) for defining non-dimensional parameters. This quantity plays a key role in the structure of the theory, as will become clear just below.

But first, to be precise and avoid ambiguities, several clarifications must be made here.

On one hand, strictly speaking one should distinguish the wave age of the actual wave from that in the airless, windless reference system. Indeed, $\theta = \theta_0 + s \, c_1/\Uwind$, from Eq.~(\ref{eq:c:expansion}). However, since $s \approx 10^{-3}$, this distinction is of no importance in practice here, and we will refer to both $\theta \approx \tho$ as 'the (monochromatic) wave age'.

On the other hand, studies of ocean waves have always associated the notion of wave age to the quantity $c_p/U_{10}$ where $c_p$ is the velocity at the wave spectrum peak, and $U_{10}$ the wind strength at a height of 10 m. This definition is clearly based on two obvious and central experimental quantities. 
However in Miles’ 57 pioneering paper, another but similar quantity appears,  $c/U$, where $c$ is the monochromatic propagation speed of the wave, which is known for a given wave vector $k$, and the wind strength $U$, which is a prefactor to the wind profile. This is the central quantity for Miles \cite{Miles:1957}, Beji and Nadaoka \cite{Beji:2004}, and for a series of articles based on Miles' approach \cite{Montalvo:2013,Montalvo:2013b,Montalvo:thesis:2014,Kern:2021,Branger:2022,Kharif:2020,Chaubet:2024}. This is also the case here. 
In reference to the 'wave age' $c_p/U_{10}$, we therefore choose to refer to the central quantity $c/U$ as the '{\it monochromatic wave age}' when we want to insist that it concerns a monochromatic wave, as opposed to a full wave spectrum.

One direct implication is that the Charnock relation, Eq.~(\ref{eq:charnock:original}), can be non-dimensionalised, multiplying by $k$, as
\begin{equation}
  \label{eq:charnock}
  \zo 
  =  \frac{\Omega_{CH}}{k \, \theta^2} 
  \ ,
\end{equation}
where  $\Omega_{CH}:=\kappa^2 \, \alpha_{CH}$.

Eq.~(\ref{eq:charnock}) is a natural generalisation of the definition in Eq.~(\ref{eq:charnock:original}), originally formulated for Miles' reference conditions, where $\theta = \theta_{dw} = \sqrt{g/k}$.

\subsection{Miles' conditions as a reference case}

Here we rewrite the approach to set it up in such a way that Miles' infinitely deep, still ocean becomes the reference case, which shows the entirely generic nature of the approach.

\paragraph{Universal role of $\Io$}

One direct benefit of this choice of non-dimensional variables becomes clear: the quantity $\Io$ (which characterises the impact of air and wind) clearly emerges as another key quantity, beyond the density ratio $s=\rho\air/\rho\wat$ and the (monochromatic) wave age $\tho = \Re[c_0]/\Uwind$. To see this, we use Eq.~(\ref{eq:Ic:def}) to express $\Io$ as
\begin{equation}
  \label{eq:I0:def}
\Io  
  :=
  \int_{\zot}^\infty \left(\frac{f(\tilde{z})}{\tho} - 1 \right) \, \frac{u_z(\tilde{z})}{u_z(\tilde\zo)} \, d(\tilde{z})
  \ .
\end{equation}
This shows clearly that $\Io$ carries no explicit dependency on the wave number $k$ nor on the characteristic wind strength $\Uwind$. Rather, it is a function of $\tho$ only (given that the non-dimensionalised roughness length $\zot$ too is commonly accepted to be a function of $\theta$ only, such as in Eq.~(\ref{eq:charnock})). This underlines further that the wave age $\tho$ clearly plays a key role in quantifying the impact of wind. It also implies that, for a given wind profile,  $\Io$ is a 'universal' quantity which depends only on the (monochromatic) wave age.

\paragraph{Characterising the wind through Miles' growth coefficient}

Using these dimensionless variables, the growth rate in Miles' reference conditions can therefore be expressed, from Eq.~(\ref{eq:gamma}), as 
\begin{equation}
  \tilde{\gamma}_M
  = \frac{k \, \Im[c_M]}{\omega_{0}}
  = \frac{1}{2} \, \Im\left[s \, \Io\right]
  \ .
    \qquad\mbox{(Miles reference conditions)}
\end{equation}

Here we use Miles' infinitely deep, still ocean as a reference system, well known in the literature. We can therefore use this relation backwards, by stating that
\begin{equation}
  \label{eq:Io:backwards}
  s \, \Im[\Io] = 2 \,  \tilde{\gamma}_M \ .
\end{equation}
Hence referring to Miles' amplitude growth coefficient $\tilde\gamma_M$ is in fact equivalent to stating the impact of wind on wave growth. Consequently, we can express the growth coefficient for specific systems either as a function of the dimensionless parameter $\Io$ or, equivalently, as a function of $\tilde\gamma_M$, which is a well established quantity from Miles' work.

\paragraph{Generalised hydrodynamic contexts}

By {\it hydrodynamic context}, we refer to the equations governing the hydrodynamics in the water domain, as well as the corresponding boundary conditions, including those at the free interface. 
For a more complex hydrodynamic context than Miles' free ocean condition discussed above, we choose to use Miles' case as a reference. To do this we compare two contributions: the dynamic contribution to the Archimedian water pressure at the interface, $\left.p\wat\right|_{z=0}$ (potentially amended by a interfacial terms) as it enters the pressure balance Eq.~(\ref{eq:pressurebalance}),  on one hand, and its equivalent for Miles' hydrodynamics, Eq.~(\ref{eq:miles:pressure}), on the other hand. We thus define the coefficient $\Pcoeff$ through
\begin{equation}
  \label{eq:hydro:Pcoeff}
  \Pcoeff
  :=
  \frac{\left.p\wat\right|_{z=0} - \Delta p}{\left.p\wat_M\right|_{z=0}}
  =
  \frac{\left.p\wat\right|_{z=0} - \Delta p}{\rho\wat k\eta \,c^2} 
  \ .
\end{equation}
Other choices are possible, but this particular coefficient leads to particularly elegant final expressions. 
Note though that $\Pcoeff$ still depends on $s$ (through $c$, which is itself a function of $s$).

A few remarks on this definition are in order.
Evidently, for Miles' conditions, this coefficient is simply $\Pcoeff_M=1$, by construction. Indeed, the coefficient $\Pcoeff$ is clearly intended to capture the extent to which the hydrodynamics differ from Miles' case, {\it at a given celerity $c$}.
To state this yet more clearly and point out its subtlety: $\Pcoeff$ quantifies the dynamic pressure which the fluid must produce to sustain the propagating wave, and it  measures it as a multiple of the pressure perturbation which the same wave, propagating  {\it with the same celerity $c$}, would produce on  Miles' ocean (static, non-viscous, infinitely deep). This may appear an awkward quantity, since there is no reason why the wave should propagate at the same speed in Miles' reference conditions (and it typically will not, since $c_M = \sqrt{g/k}$ is a specific value).
Nevertheless, it will become clear below that this is indeed a convenient definition.

\paragraph{Generalised dispersion relation}

Using Eq.~(\ref{eq:hydro:Pcoeff}) the pressure balance at the interface, which has lead to the dispersion equation Eq.~(\ref{eq:dispersion}), can now be written, for a general hydrodynamic context,  as
\begin{equation}
  \label{eq:dispersion:full}
  (1-s) = \Pcoeff \, \frac{c^2}{c_{M,0}^2} - s \, \frac{c^2}{c_{M,0}^2} \, \Io
  \ .
\end{equation}
Recall that the coefficient $\Pcoeff$ characterises the hydrodynamics in the water domain, whereas the coefficient $\Io$ defined in Eq.~(\ref{eq:I0:def}) captures the impact of air (including wind).
Recall also that determining $\Io$ requires solving the Rayleigh equation (\ref{eq:Rayleigh}), but is readily available from Miles' reference conditions alone.

The dispersion equation (\ref{eq:dispersion:full}) is the point which all further considerations will hinge upon.

\paragraph{Dissipation: handling imaginary contributions to the celerity}

The celerity $c_0$, in the absence of air and wind, may have an imaginary part, which we have not explicitly acknowledged so far. This consideration turns out to be an aside, which adds no further complication, but it must be addressed in order to make our argument general.

Note therefore that an imaginary part $\Im[c_0]$, as it would arise from dissipation, is seen from Eq.~(\ref{eq:gamma}) to result in a non-zero amplitude growth coefficient even in the absence of wind. Which is of course expected to be negative, reflecting a decay in amplitude due to dissipative processes.
However, in the present approach it is easy to see that this is not important as long as the imaginary contribution to $c_0$ is 'small', a fact already illustrated numerically by Stiassnie {\it et al.} \cite{Stiassnie:2007}. To see this, recall the dispersion relation Eq.~(\ref{eq:dispersion:full}), in which the wind interferes only through the term $s \times \Icoeff$. From the definition of $\Icoeff$, Eq.~(\ref{eq:I0:def}), it is clear that, just as replacing $c\approx c_0$ in the integral neglects contributions of order $\oforder{s^2}$, using only the real part of the celerity  will neglect terms of order $\oforder{s \Im[c_0]/\Re[c_0]}$. As long as these may be assumed to be small, it is therefore entirely sufficient to work with the real part of $c_0$, taking $c_0 \approx\Re[c_0]$, in the associated Rayleigh Eq.~(\ref{eq:Rayleigh}).

\subsection{Generalising beyond Miles' conditions}

In essence, recapitulating Miles' approach (in its initial context) has exposed the following points.
  
  (i) the presence of air (and therefore, a fortiori, of wind) impacts on the quantity $\Icoeff$ {\it only}, which is furthermore independent of all other considerations (hydrodynamics in the water domain, interfacial effects, etc).

  (ii) the dispersion relation is dependent on air (and wind) through the quantity $s \times \Icoeff$ {\it only}, and at the level of this equation, the impact of air and wind is therefore {\it independent} of the precise hydrodynamic context in the water domain or interfacial effects.

  (iii) hydrodynamics and interfacial effects enter the dispersion relation through the term $\Pcoeff$ {\it only}~: it is therefore sufficient to know the pressure perturbation due to the passing wave in the water domain, as well as any interfacial contributions, in order to account for the impact of hydrodynamics in the water domain.

We will now show formally that, in order to determine the wave growth rates  in a general hydrodynamic context, it is in fact sufficient to know the water pressure perturbation at $s=0$, \ie in an airless, windless system.

\paragraph{Solving for the celerity}

Evaluating the generalised dispersion relation, Eq.~(\ref{eq:dispersion:full}), for $s=0$ yields
\begin{equation}
    \label{eq:c0}
  c_0^2 = \frac{c_{M,0}^2}{\Pcoeff_0} 
\end{equation}
from which we can rewrite the dispersion relation as
\begin{equation}
  c_0^2 \,  (1-s) = c^2 \, \left(\frac{\Pcoeff}{\Pcoeff_0} - s \, \frac{\Io}{\Pcoeff_0}\right)
  \ .
\end{equation}
Expanding this in powers of $s$, using $\Pcoeff = \Pcoeff_0 + s \, \Pcoeff_1$ and $c = c_0 + s \, c_1$ and comparing the linear coefficients leads to an expression for the first-order correction to the celerity $c_1$:
\begin{equation}
  \label{eq:c1/c0:rewrite}
  \frac{c_1}{c_0}
  =  \frac{1}{2} \, \left(\frac{\Io}{\Pcoeff_0}  - 1 - \frac{\Pcoeff_1}{\Pcoeff_0}\right)
  \ .
\end{equation}
In the case where $\Pcoeff_1=0$, this directly provides the full celerity $c=c_0 (1+s \, c_1/c_0)$, and therefore the growth coefficient $\tilde\gamma$, via Eq.~(\ref{eq:gamma}). We will see in the examples below that this is often sufficient. However, in the general case $\Pcoeff_1 \neq 0$, and the analysis needs to be pursued further.

\paragraph{Determining the pressure coefficient $\Pcoeff$}

Taylor-expanding the pressure condition as $\Pcoeff = \Pcoeff_0 + s \, \Pcoeff_1 + \oforder{s^2}$ we have
\begin{equation}
  \Pcoeff_1
  = \left.\der{\Pcoeff}{s}\right|_{s=0} 
  = \left.\pder{\Pcoeff}{c}\right|_{s=0} \, \left.\der{c}{s}\right|_{s=0}
  \ ,
\end{equation}
and consequently 
\begin{equation}
  \label{eq:Phi:expansion}
  \Pcoeff_1
    = \left.\pder{\Pcoeff}{c}\right|_{s=0}  c_1
  \ .      
\end{equation}

This is an important observation, since it establishes that the first-order correction due to the presence of air $(s\neq0$), specifically $\Pcoeff_1$ and $c_1$,  are linked by a coefficient based on an analysis at $s=0$, \ie which does not require dealing with the air (and, a fortiori, with wind). This is ultimately what makes the definition used in Eq.~(\ref{eq:hydro:Pcoeff}) a strategic choice.

\paragraph{General expression for the amplitude growth rate}

Closure is now achieved by substituting $\Pcoeff_1$ from Eq.~(\ref{eq:Phi:expansion}) into the full dispersion relation expressed as in Eq.~(\ref{eq:c1/c0:rewrite}) and solving for $c_1/c_0$. The result can be written as
\begin{equation}
  \label{eq:c}
  c = c_0 \, \left(1 + s \, \Upsilon_0\right) + \oforder{s^2}
  \ ,
\end{equation}
where we have introduced the convenient coefficient
\begin{equation}
  \label{eq:Upsilon0:def}
  \Upsilon_0
  :=
  \frac{c_1}{c_0}
  =  \frac{\frac{1}{2}  \left(\frac{\Io}{\Pcoeff_0} - 1\right)}{1 + \frac{1}{2}\frac{c_0}{\Pcoeff_0} \pder{\Pcoeff_0}{c_0}}
  \ .
\end{equation}

Importantly, $\Upsilon_0$ can be calculated in a very straightforward way knowing only the expression for the pressure coefficient $\Pcoeff_0$, \ie without the need to couple the water hydrodynamics to the air flow. Note also its value in Miles' conditions, $\Upsilon_{0,M} = \frac{1}{2} \, (\Io-1)$, from which we correctly recover Miles' result Eq.~(\ref{eq:cM1}) for $c_{M,0}$ based on Eq.~(\ref{eq:c}).
\\

Since the amplitude growth coefficient $\gamma$ follows directly from its definition Eq.~(\ref{eq:gamma}), it is clear from Eqs.~(\ref{eq:c}) and (\ref{eq:Upsilon0:def}) that we have reached the pivotal point of Miles' approach, which can now be formulated in a more general way: the effect of wind on wave growth can be determined based solely on the coefficient $\Pcoeff_0$ (which characterises the hydrodynamic situation \textit{in the absence of air and wind}, see Eq.~(\ref{eq:hydro:Pcoeff})), and on the integral $\Io$ (which entirely captures the interaction with air flow, and wind).

\subsection{Generalised amplitude growth coefficient $\gamma$}

We can now establish the central result,  a simple and easy to use expression for the amplitude growth coefficient in general hydrodynamic conditions.

\paragraph{An intermediate result}
The non-dimensional amplitude growth coefficient $\tilde\gamma = \gamma/\omega_0$ now follows directly from its definition, Eq.~(\ref{eq:gamma}), and Eq.~(\ref{eq:c}), as
\begin{equation}
  \label{eq:gammatilde:expression}
  \tilde\gamma 
  = \frac{\Im\left[ c_0 \times (1 + s \, \Upsilon_0)\right]}{\Re[c_0]} 
  \ .
\end{equation}
We simplify this expression further before exploiting it.

\paragraph{Simple case of real $c_0$}
When there is no dissipation in the fluid, $c_0$ is real, and so is $\Pcoeff_0 = c_{0,M}^2/c_0^2$. Therefore Eq.~(\ref{eq:gammatilde:expression}) simplifies to $\tilde\gamma = s \, \Im[\Upsilon_0]$ in this case. Substituting Eq.~(\ref{eq:Upsilon0:def}) into this and taking the imaginary part thus yields
\begin{equation}
  \label{eq:gammatilde:c0real}
  \tilde\gamma
  = \Im[s \, \Upsilon_0]
  = \frac{s}{2} \, \frac{1}{\Pcoeff_0} \, \frac{\Im[\Io]}{1 + \frac{1}{2}\frac{c_0}{\Pcoeff_0}  \left.\pder{\Pcoeff_0}{c_0}\right.
    }
  =
  \Xcoeff_0
    \times \tilde\gamma_M
  \qquad\mbox{($c_0$ real)}
  \ ,
\end{equation}
where
\begin{equation}
  \label{eq:chi0:def}
  \Xcoeff_0
  :=
  \frac{1}{\Pcoeff_0} \, \frac{1}{1 + \frac{1}{2} \frac{c_0}{\Pcoeff_0}  \pder{\Pcoeff_0}{c_0}}
\ .
\end{equation}
Therefore, the growth rate can easily be calculated, and is in fact simply the growth coefficient $\tilde\gamma_{M}$ in Mile's conditions rescaled by a factor $\Xcoeff_0$ which characterises the hydrodynamics.

\paragraph{Dissipation: handling imaginary contributions to the hydrodynamics}

When $c_0$ includes an imaginary contribution, further consideration is required to extend the discussion initiated above. 
In this case, we can obtain the amplitude growth rate by adding two contributions,
\begin{equation}
  \label{eq:gamma:viscosity:additivity}
  \tilde\gamma = \left.\tilde\gamma\right|_{s=0}  + \left.\tilde\gamma\right|_{\Im[c]=0}
  =\tilde\gamma_{diss} +\tilde \gamma_{wind}
\end{equation}
Although a detailed mathematical demonstration (which we provide in Appendix~\ref{app:dissipation}) is of course reassuring, the origin of this additivity is in fact straightforward: since we have neglected second order effects of both the impact of air ($s\ll 1$) and of dissipation ($\Im[c_0]/\Re[c_0] \ll 1$), there can be no combined effect, as this would also be of second order.

Therefore the (positive) growth rate due to wind and the (negative) contribution from dissipation in the water domain are additive. Just as the latter,
\begin{equation}
  \label{eq:def:gamma:diss}
  \tilde\gamma_{diss} :=  \left.\tilde\gamma\right|_{s=0} = \left.\frac{\Im[c]}{\Re[c]}\right|_{s=0}
  \qquad\mbox{\ie}\qquad
  \tilde\gamma_{diss}=\tilde\gamma_0  = \frac{\Im[c_0]}{\Re[c_0]}
    \ ,
\end{equation}
is evaluated without air ($s=0$), the former is evaluated without dissipation ($\Im[c_0]=0)$, as
\begin{equation}
    \tilde\gamma_{wind} \approx s \, \Im\left[\left.\Upsilon_0\right|_{\Im[c_0]=0}\right]
  \ .
\end{equation}
Essentially, the impact of wind is just what it would be without dissipation, and can therefore be calculated from Eq.~(\ref{eq:gammatilde:c0real}), simply replacing $\Pcoeff_0$ and $c_0$ by their real parts, \ie one may take
\begin{equation}
    \label{eq:gammawind:approx}
    \Xcoeff_0 \approx \left.\Xcoeff_0\right|_{\Im[\Pcoeff_0]=0}
    \qquad\mbox{and}\qquad
    \tilde\gamma_{wind} \approx \Xcoeff_0 \times \tilde\gamma_M
\end{equation}
to leading order.

\section{Beyond Miles: applying the framework in  practice\label{sec:examples}}

We now provide examples to illustrate just how little algebra is required in order to derive the growth coefficient from the  expression for $\Pcoeff_0$, which characterises the hydrodynamics in the water domain.
To do so, we apply the procedure established above to cases for which the expected results are already known in the literature. The first two examples do no  involve interfacial stresses, such as those due to a surface tension, and so we  take $\Delta p=0$. The last example will require considering $\Delta p$ due to viscous stresses.

\subsection{Example: finite water depth}

As a straightforward example, we turn to a situation of finite depth without currents. Then the dynamic water pressure for a wave propagating over the flat interface, with $s=0$, is well known to be \cite{Janssen:book,Kinsman:book}
\begin{equation}
  p\wat(z) = k \rho\wat c^2 \, \frac{\cosh(k(z+h))}{\sinh(kh)} \, \eta \, e^{i\varphi}
  \ ,
\end{equation}
in terms of a normal mode analysis (as in section~\ref{sec:essence}), where $h$ ist the water depth.
Evaluated at $z=0$, this yields the leading order (since the air has not been considered at this stage) to Eq.~(\ref{eq:hydro:Pcoeff}) as
\begin{equation}
  \Pcoeff_0=1/\tanh(\tilde{h})
  \ .
\end{equation}

Furthermore, since $\pder{\Pcoeff_0}{c_0}=0$, we have from Eq.~(\ref{eq:chi0:def}) that
\begin{equation}
  \Xcoeff_0 = 1/\Pcoeff_0 = \tanh(\tilde{h})
  \qquad\mbox{and thus}\qquad
  \tilde\gamma = \tanh(kh) \times \tilde\gamma_M
\end{equation}
without any need for tedious calculations. The result is as found in \cite{Montalvo:2013}.

This is in fact a special case of the more complete example we address now.

\subsection{Example: underwater shear gradient}

When a constant shear gradient is present in the fluid, the imposed flow field for the water can be expressed in terms of the (constant) vorticity $\Omega$ as
\begin{equation}
  V_{ext} = \Omega \, z
  \ .
\end{equation}
For this case the expression for the dynamic water pressure perturbation of a wave propagating at celerity $c$  is known to be (Eq.~(3.1) in \cite{Hur:2019}, 
\footnote{Once correspondence is made:  relation (3.1) in \cite{Hur:2019} omits the prefactor in $\rho\wat \times \eta$, which is obvious on dimensional grounds; also, the (constant) vorticity is noted $\omega$, which is the negative of our $\Omega$.}
%
\begin{equation}
    \displaystyle p\wat = \displaystyle \frac{\rho\wat k c \, \eta}{\sinh(kh)} \, \big[ (c-\Omega \, z) \, \cosh(k(z+h))
       \displaystyle + \frac{\Omega}{k} \, \sinh(k(z+h))\big]
    \ .
\end{equation}
%
Evaluating this at $z=0$ yields the hydrodynamic coefficient as
\begin{equation}
  \Pcoeff = 1/\tilde{T} + \tilde{\Omega}\,{c_0/c} 
  \qquad\mbox{and thus}\qquad
  \Pcoeff_0 = 1/\tilde{T} + \tilde{\Omega}
  \ ,
\end{equation}
where we have introduced the non-dimensional hydrodynamic parameters
\begin{equation}
  \tilde{T} := \tanh(kh)
  \qquad\mbox{and}\qquad
  \tilde{\Omega} := \frac{\Omega}{\omega_0}  = \frac{\Omega}{k \, c_0}
  \ .
\end{equation}

Here, $\Pcoeff_0$ does depend on $c_0$, and one finds without difficulty that
\begin{equation}
  \frac{c_0}{\Pcoeff_0}\pder{\Pcoeff_0}{c_0} = - \frac{\tilde{\Omega}}{\Pcoeff_0}
  \ ,
\end{equation}
from which we have directly, via Eq.~(\ref{eq:chi0:def}), that
\begin{equation}
  \label{eq:gammatilde:shearflow}
  \Xcoeff_0 = \frac{1}{\Pcoeff_0 -  \tilde{\Omega}/2} = \frac{1}{1/\tilde{T} + \tilde\Omega/2}
  \qquad\mbox{and thus}\qquad
  \tilde\gamma = \frac{\tilde{T}}{1+\tilde{\Omega}\tilde{T}/2} \times \tilde{\gamma}_M
  \ .
\end{equation}
This is in agreement with \cite{Kern:2021}, albeit expressed in a different form (see the Appendix \ref{app:previous} for full details of this correspondence).

\subsection{Example: water with a bulk viscosity}

The effect of viscosity in the water domain has received considerable attention in experiments \cite{Paquier:PhD,Paquier:2015,Paquier:2016,Geva:2022,Zhang:2023}, and has been studied very recently theoretically \cite{Wu:2021,Wu:2022,Chaubet:2024}, using a complete derivation to generalise the calculations proposed by Miles. We show here, for deep water conditions, that our approach delivers the result rather easily, based on the pressure at the interface, in a reference system with a constant homogeneous air pressure, through straightforward algebra.

Again, we start from the expression for the dynamic pressure perturbation in an airless system. It has been given by Dias {\it et al.} \cite{Dias:2008} as $i\, 2 \rho \eta k^2 \nu c$ (see their Eq.~(24)), to be complemented by the viscous interfacial stress contribution (their Eq.~(6)), which ultimately leads to
\begin{equation}
  \left.\bigg( \left.p\wat\right|_{s=0} - \left.\Delta p\right|_{s=0} \bigg)\right|_{z=0} = i \, 4 \rho \eta k^2 \nu c
  \ .
\end{equation}
Different methods for calculating the pressure have been reconciled by Eeltink et al. \cite{Eeltink:2020}, who give the same result, which is furthermore in agreement with the derivation in \cite{Chaubet:2024} when taking $s=0$.

Therefore, defining the non-dimensional viscosity
as $\tilde\nu:=k\nu/\Re[c_0]$,
the pressure coefficient is
\begin{equation}
  \label{eq:Phi0:viscosity}
  \Pcoeff_0 = 1 + i \, 4 \tilde{\nu}
  \ ,
\end{equation}
but according to Eq.~(\ref{eq:gammawind:approx}) the imaginary contribution is to be neglected when calculating the impact of wind, which simply means $\Pcoeff_0 \approx 1$ to leading order. This just states that (i) to leading order, there is no impact of viscosity on the energy transfer from wind to the waves, as already discussed above, and (ii) there are no other hydrodynamic effects since, except for the presence of dissipation, the hydrodynamic conditions are just those of Miles. Therefore
\begin{equation}
  \Xcoeff_0 \approx 1
  \qquad\mbox{and thus}\qquad
  \tilde{\gamma}_{wind} = \tilde{\gamma}_{M}
  \ ,
\end{equation}
From Eqs.~(\ref{eq:gamma:viscosity:additivity}) ans (\ref{eq:c0}) we have
\begin{equation}
  \tilde\gamma \approx \tilde\gamma_M  - 2 \tilde\nu
  \ ,
\end{equation}
which reflects the additivity between viscous damping and wind-driven wave growth. This is indeed the result obtained in \cite{Chaubet:2024} once conversion is made between the non-dimensional variables used here and there.

\section{Beyond Miles: propagation speed and pressure coefficients}\label{sec:others}

Although the amplitude growth rate $\tilde\gamma$ which we have focused on  so far is clearly a central quantity, there are other parameters which are useful to determine. Here we give results for the propagation speed of the wave, and for a generalisation of Miles' pressure coefficient (usually referred to as $\beta$), which is also widely used in order to characterise wave growth.

\subsection{Zero-wind limit {\it vs.} 'airless' limit}

Before doing so, in order to avoid confusion when interpreting results, we specifically mention the distinction between the limits of vanishing wind {\it vs.} the limit where the air is totally absent (replaced by a constant pressure field, such as for a perfect gas, once the weight of the air molecules is taken to vanish).

Let us thus consider the limit of vanishing wind ($\Uwind=0$) while maintaining the weight of the air ($s \neq 0$). In this case
\begin{equation}
  \label{eq:I0:s=0}
  \Io=-1 \qquad\mbox{for}\qquad  U_{ext}=0
    \ ,
\end{equation}
which is easily established by noting that the air pressure perturbation, according to the Rayleigh equation Eq.~(\ref{eq:Rayleigh}), directly leads to an exponentially decaying velocity profile $u_z \sim \exp(-kz)$ in the air domain. From this, the integral $\Io$ follows directly as stated.

Clearly, $\Icoeff_0$ does not vanish in the absence of wind, and not even in the absence of air. This is because $\Icoeff_0$, defined in Eq.~(\ref{eq:Ic:def}), characterises the impact of air flow as a whole.
It thus covers two aspects: the explicit appearance of the wind profile $U_{ext}$ in the integrand, and its implicit impact, through Eq.~(\ref{eq:Rayleigh}), on the air flow $u_z(z)$ accompanying the propagating wave. In a sense, for  the  'windless' limit ($U_{ext} \to 0$), $\Io$ thus characterises the air circulation accompanying the passing wave. In the 'airless' limit ($s \to 0$), although the quantity $\Io$ remains well defined, and non-zero, it nevertheless produces no wave growth, since it carries a prefactor $s$ (see Eqs.~(\ref{eq:Upsilon0:def})and (\ref{eq:gammatilde:expression})).

\subsection{Propagation speed and wave age}

The impact of air and wind on the propagation speed of the wave is often not explicitly determined, but is so easily available from our previous results in a generic way that we take the detour to state the result here.

A generic result can be established by taking the real part of Eq.~(\ref{eq:c}):
\begin{equation}
  \Re[c]
  =  \Re[c_0]   + s \, \Re[c_0] \, \Re[\Upsilon_0] - s \, \Im[c_0] \, \Im[\Upsilon_0]
  \ .
\end{equation}
We can now proceed as we have done for the imaginary part of this quantity: to leading order, dropping cross-terms between $s$ and $\Im[c_0]/\Re[c_0]$, the last term is negligible.
Thus the propagation speed, including its first-order correction, is
\begin{equation}
  \label{eq:celerity:dissipation}
  \Re[c] = \Re[c_0] \, \left( 1 + s \, \Re[\Upsilon_0] \right)
    \ .
\end{equation}

Note that a special case of  gravity waves can easily be recovered for the windless ($\Io=-1$) limit of Miles hydrodynamics ($\Pcoeff_0=1$). In this case Eq.~(\ref{eq:Upsilon0:def}) reduces to $\Upsilon_0 = -1$, which implies from Eq.~(\ref{eq:celerity:dissipation}) that $c = (1-s)\, \sqrt{g/k}$. This is the first-order correction due to a factor $\sqrt{(1-s)/(1+s)}$ which is indeed encountered when taking into account the presence of air in gravity waves \cite{Kinsman:book,Landau:book,Lamb:book}.

Note that this coefficient also applies to the (monochromatic)wave age,
\begin{equation}
 \theta = \tho \, \left( 1 + s \, \Re[\Upsilon_0] \right)
\end{equation}
and therefore one should conceptually distinguish $\theta$ (as it is measured, in the presence of air and wind) from the reference wave age $\tho$ (which is required for calculations). As already stated, however, the correction is entirely negligible (or at least so for an ocean of water under an atmosphere of air on earth).

\subsection{Miles' pressure coefficients}

So far our analysis has focused on the amplitude growth coefficient $\tilde\gamma$.
However, another growth coefficient has been introduced by Miles, which is widely used and also particularly interesting: it is directly related to the water pressure at the interface, which we have put at the centre of our argument.

Indeed, Miles \cite{Miles:1957} chooses to define a complex coefficient, $\alpha+i \beta$, with real $\alpha$ and $\beta$, as
\begin{equation}
  \label{eq:pressurecoeffs:Miles}
  \left.p\right|_{z=\eta(x,t)} = (\alpha + i \beta) \, \rho k \eta \, \Uwind^2 
  \qquad\mbox{(Miles reference conditions)}
  \ ,
\end{equation}
which Miles evaluates explicitly in the airless system ($s=0$).
These (real) coefficients are particularly convenient, since they characterise the phase lag between the wave and the pressure perturbation, and thus ultimately the energy transfer to the wave.

This definition has the advantage of being defined exclusively in the air domain. We therefore generalise the definition, based on Eq.~(\ref{eq:pressure:air:interface}), as
\begin{equation}
  \alpha_0 + i \beta_0 := \left[\frac{\Re[c]^2}{\Uwind^2} \, \Icoeff\right]_{s=0}
  \ .
\end{equation}
so that 
\begin{eqnarray}
  \alpha_0 &=& \tho^2 \, \Re[\Io] \equiv \alpha_M
  \\
  \beta_0 &=& \tho^2 \, \Im[\Io] \equiv \beta_M
  \label{eq:beta}
  \ .
\end{eqnarray}
This essentially extends Miles' definition as it is. $\alpha_0$ and $\beta_0$ therefore depend neither on $s$ (since they are defined at $s=0$ or, more precisely, for weightless air, $\rho\air \to 0$), nor on the hydrodynamic context (since, at given $c_0$, its definition refers only to the air domain).
\\

In practice, rather than working with the amplitude growth coefficient $\tilde\gamma$, one may thus prefer to use Miles' pressure coefficient $\beta_M$, which is well studied and partially tabulated, for example in \cite{Beji:2004}.
Using Eqs.~(\ref{eq:gammatilde:c0real}), (\ref{eq:Io:backwards}) and (\ref{eq:beta}), we then have $\tilde\gamma$ from
\begin{equation}
  \tilde\gamma
  = \Xcoeff_0 \times \tilde\gamma_M
  =
  \Xcoeff_0
  \times \frac{s}{2}  \times \frac{\beta_M}{\tho^2}
  \qquad\mbox{with}\qquad
  \Xcoeff_0=  \frac{1}{\Re[\Pcoeff_0]} \, \frac{1}{1 + \frac{1}{2} \frac{\Re[c_0]}{\Re[\Pcoeff_0]}  \pder{\Re[\Pcoeff_0]}{\Re[c_0]}}
  \ ,
\end{equation}
which thus separates the effect of wind-driven wave growth into three multiplicative contributions: hydrodynamics (easily evaluated as described here), the air density (via $s=\rho\air/\rho\wat$), and the effect of air flow and wind.

It is thus possible to exploit the coefficient $\beta_M$, which is more commonly used and more readily available, in order to deduce the amplitude growth rate in hydrodynamic conditions other than the simple reference conditions introduced by Miles.

\subsection{Putting the celerity at the centre of the argument}
This paragraph essentially produces expressions for the growth coefficient $\tilde{\gamma}$ deriving it directly from the expression for the celerity $c_0$ and its dependency on the hydrodynamic conditions, rather than from the pressure coefficient $\Pcoeff_0$.
This is indeed potentially simpler in practice, since an expression for the celerity is often more readily available in the literature than the one for the pressure perturbation.

\paragraph{Functional form of the celerity}
The relationship between the phase velocity $c_0$ and that in Miles' system, $c_{0,M}$, may be expressed formally as
\begin{equation}
  c_0 = F_0(c_{0,M},\tilde{h}_1,\tilde{h}_2,...)
  \ ,
\end{equation}
where the variables $\tilde{h}_i$ stand for all (non-dimensional) hydrodynamic parameters required in the situation of interest. This indicates that, beyond  $c_{0,M}=\sqrt{g/k}$,  the celerity is set by the hydrodynamics. This identity implies that $c_0$, $c_{0,M}$ and all the $\tilde{h}_i$ are (of course) not independent parameters. But note also that the $\tilde{h}_i$ may depend on $c_0$, as is the case when there is a hydrodynamic parameter which requires a velocity scale ($c_0$)  or a time scale ($k c_0$) for non-dimensionalisation. Therefore this is in a sense a self-consistent equation.

It is now tempting to think that there should be a factorisation, \ie
\begin{equation}
  F_0(c_{0,M}, \tilde{h}_1,\tilde{h}_2,...) = c_{0,M} \times \Hcoeff_0(\tilde{h}_1,\tilde{h}_2,...) 
  \,
\end{equation}
with some adequate function $\Hcoeff_0(\tilde{h}_1,\tilde{h}_2,...)$. And this is indeed what we have already shown, see Eq.~(\ref{eq:c0}), which essentially states that
\begin{equation}
  \label{eq:f:def}
  \Hcoeff_0(\tilde{h}_1,\tilde{h}_2,...) = \frac{1}{\sqrt{\Pcoeff_0}} = \frac{c_0}{c_{0,M}}
  \ .
\end{equation}
This now opens up two routes to proceed. First though we note in passing that the choice of the symbol $\Hcoeff$ is deliberate, and is meant to stand for 'hydrodynamics': we shall show in the following that this coefficient indeed captures
the impact of hydrodynamics completely.

\paragraph{Deducing the pressure perturbation from the celerity}

One route is to follow the approach as outlined above, based on the pressure coefficient $\Pcoeff_0$, which can thus be deduced from the expression for the celerity in the airless system.

Remarkably, this also implies that
\begin{equation}
  p = \Pcoeff_0 \times p_M = \frac{1}{\Hcoeff_0^2} \times p_M
  \ ,
\end{equation}
\ie the amplitude of the pressure perturbation is that of Miles' system, rescaled by a coefficient which is simply the square of the ratio $\Hcoeff_0 = c_0/c_{0,M}$, which expresses the impact of the hydrodynamics on the celerity.

\paragraph{Working directly from the celerity}
The other route is to completely bypass the pressure coefficient in terms of the ratio of celerities defining $\Hcoeff_0$. 
To this end, observe first that in the expression of $\Xcoeff_0$, Eq.~(\ref{eq:chi0:def}), we have
\begin{equation}
  \pder{\Pcoeff_0}{c_0}
  = \pder{}{c_0} \left(\frac{1}{\Hcoeff_0^2(\tilde{h}_1,\tilde{h}_2,...)}\right)
  = - \frac{2}{\Hcoeff_0^3} \, \pder{\Hcoeff_0(\tilde{h}_1,\tilde{h}_2,...)}{c_0}
  \ ,
\end{equation}
where the derivative really stands for a sum over all hydrodynamic parameters which carry a dependency on $c_0$:
\begin{equation}
  \label{eq:Ho:pder}
  \pder{\Hcoeff_0(\tilde{h}_1,\tilde{h}_2,...)}{c_0}
  =
  \sum_i \pder{\Hcoeff_0(\tilde{h}_1,\tilde{h}_2,...)}{\tilde{h}_i} \, \pder{\tilde{h}_i}{c_0}
  \ ,
\end{equation}
although these terms do not contribute to the sum whenever $\tilde{h}_i$ does not require $c_0$ for non-dimensionalisation.

Substituting this, as well as the definition of the function $\Hcoeff_0(...)$, we can thus express Eq.~(\ref{eq:chi0:def}) as
\begin{equation}
  \label{eq:chi0:f}
 \Xcoeff_0
 =  \frac{\Hcoeff_0^2}{1 - \frac{c_0}{\Hcoeff_0} \, \pder{\Hcoeff_0}{c_0}}
 =  \frac{\Hcoeff_0^2}{1 - c_{0,M} \, \pder{\Hcoeff_0}{c_0}}
  \ .
\end{equation}

It is therefore, via Eq.~(\ref{eq:f:def}), possible to work directly from 
celerity $c_0$ in the absence of air (and wind).

\section{Summary, discussion and outlook} \label{sec:discussion}

The aim of this manuscript has been to facilitate extending Miles' first-principles approach for wind-wave interaction to hydrodynamic conditions beyond those initially envisaged by Miles, \ie a still, infinitely deep ocean. We have reviewed the essence of Miles' approach, short of reproducing a full account of all derivations, but with the intention of exposing the major arguments.
A slight change of focus, putting the pressure balance at the water-air interface at the center of the argument, is key to the following generalisations: handling the two media, air and water, separately, before coupling them  through the pressure at the interface on which the wave propagates, exposes that this procedure is in fact generic. 

Maintaining Miles' analysis of the air flow we have then shown that the coupling to the hydrodynamics in the water domain, whatever its precise hydrodynamic origin, can be characterised in terms of a single coefficient, $\Pcoeff_0$. It relates the dynamic water pressure perturbation which the wave produces in the water domain to the pressure the same wave would produce in Miles' conditions (adding, if required, interfacial terms).
Importantly, to determine this coefficient it is sufficient to solve the water hydrodynamics in a simplified system, without air (and thus without wind), which is a significant simplification.

The impact of the energy transfer from wind to wave can then be determined from this  coefficient alone, via a simple algebraic calculation (see Eq.~(\ref{eq:chi0:def})). It results in a hydrodynamic factor, $\Xcoeff_0$, by which Miles' amplitude growth coefficient $\tilde\gamma_M$ must be rescaled.  In a sense one could say~-~somewhat provocatively~-~that amplitude growth coefficient contains for Miles' idealised reference case is in fact all there is to know about energy transfer from wind to waves.
We have demonstrated, through several examples, that this correctly reproduces the results obtained before by a complete analysis.

Furthermore, we have then exploited the fact that the amplitude of the pressure perturbation follows directly from the dispersion relation.
This yields a further shortcut, providing the hydrodynamic coefficient for wave growth directly from the expression for the wave celerity of a water wave in the single-component system, where the air is replaced by a constant pressure field.

In formulating this framework, we have also extracted several generic results.
For example, although dissipative processes in the water domain raise important questions, we have shown that when dissipation is weak, one obtains two separate growth coefficients to wind (positive) and to dissipation (negative). To leading order, these contributions are additive, and  the contribution due to wind can be established by simply ignoring any dissipation.

In summary, our work provides a generic and easy to use formulation for extending Miles' approach for wind-wave interaction, with comparatively little effort, to much more complex hydrodynamic scenarios. It neither requires additional numerical work nor does it introduce any adjustable parameters. Future generalisations should include, amongst others, the role of surface tension and the role of bottom friction.
We hope that this framework will help to make this kind of analysis tractable, and therefore usable in practice, for many future studies which may contribute to deepening our understanding of ocean waves.

\subsection*{Acknowledgements} 
We wish to thank  F. Bouchette, H. Branger, J. Dorignac and F. Geniet for many useful interactions.
We acknowledge funding via the program 'France 2030' under the reference ‘ANR-23-EXMA-0007'.

\subsection*{Declaration of Interests}
The authors report no conflict of interest.

\bibliography{framework}
\bibliographystyle{unsrt}

\clearpage

\appendix

\section{Growth rates in the presence of dissipation}
\label{app:dissipation}

In the main text we have argued that, when evaluating the growth rate due to air and wind from the hydrodynamic coefficient $\Pcoeff_0$ following Eq.~(\ref{eq:gammatilde:expression}), it is sufficient to consider the real part $\Re[\Pcoeff_0]$ only. 

In order to demonstrate this formally, the simplest way is to proceed from Eq.~(\ref{eq:chi0:f}), which uses the function $\Hcoeff_0$ characterising the ratio of celerities, rather than the ratio of pressures expressed through $\Pcoeff_0$. We thus have
\begin{equation}
  s \times \Xcoeff_0 = s \times \frac{\Hcoeff_0^2}{1-c_{0,M} \, \pder{\Hcoeff_0}{c_0}}
  \ ,
\end{equation}
where we keep the explicit prefactor $s \times$ as a reminder that this quantity always intervenes multiplied by $s \ll 1$ (or by $\tilde\gamma_M$, which itself carries a factor $s$).

Decomposing into real and imaginary contributions we can thus write
\begin{eqnarray}
  s \times \Xcoeff_0
  &=&
  s \times  \frac{\Hcoeff_0^2}{1- c_{0,M} \pder{\Re[\Hcoeff_0]}{\Re[c_0]} - i \,  c_{0,M} \pder{\Im[\Hcoeff_0]}{\Re[c_0]}}
  \\
  &\approx& s \times
  \frac{\left(\Re[\Hcoeff_0]+ i \, \Im[\Hcoeff_0]\right)^2}{1- c_{0,M} \pder{\Re[\Hcoeff_0]}{\Re[c_0]}}
  \times \left(1+ i \, \frac{ c_{0,M} \pder{\Im[\Hcoeff_0]}{\Re[c_0]} }{1- c_{0,M} \pder{\Re[\Hcoeff_0]}{\Re[c_0]} }\right)
  \ ,
\end{eqnarray}
where we have also assumed $\Hcoeff(..)$ to be an analytic function, such that the derivative in $c_0$ may simply be taken in the real direction. Also, $c_{0,M}=\sqrt{g/k}$ is real.

Since dissipation is assumed to be weak, we have $\Im[\Hcoeff_0] \ll \Re[\Hcoeff_0]$, and we can neglect terms of order $s \times \Im[\Hcoeff_0] / \Re[\Hcoeff_0]$, as well as the mixed product between $s$ and the  imaginary contribution in the parenthesis. This therefore leads to assuming
\begin{equation}
  s \times \Xcoeff_0 = s \times \frac{\left(\Re[\Hcoeff_0]\right)^2}{1- c_{0,M} \pder{\Re[\Hcoeff_0]}{\Re[c_0]}}
  = s \times \left.\frac{\Hcoeff_0^2}{1- c_{0,M} \pder{\Hcoeff_0}{c_0}}\right|_{\Im[c_0]=0}
  \ .
\end{equation}
We have thus neglected terms of order $\oforder{s^2}$ and $\oforder{s \times \Im[c_0]/\Re[c_0]}$, in keeping with our previous assumptions, and we indeed recover the same expression as for real $c_0$, Eq.~(\ref{eq:gammatilde:c0real}), where both $\Pcoeff_0$ and $c_0$ have been replaced by their real parts: 
\begin{equation}
  s \times \Xcoeff_0
   \approx
   s \times \left.\Xcoeff_0\right|_{\Im[c_0]=0}
   \qquad\mbox{and thus}\qquad
   \tilde{\gamma}_{wind} =  \left.\Xcoeff_0\right|_{\Im[c_0]=0} \times \tilde\gamma_M
  \ .
\end{equation}
This proves our point.

\section{Correspondence with previous results}
\label{app:previous}

The following is a complete demonstration that the results established here, for a finite water depth and constant vorticity shear currents, are entirely equivalent to those we have obtained before, elsewhere \cite{Kern:2021}, through a full derivation, considering the coupling to the air domain throughout the complete analysis of the hydrodynamics.

We start from the expression for $\tilde\gamma$ obtained in \cite{Kern:2021}, which, from Eq.~(56) in \cite{Kern:2021}, is
\begin{equation}
  \label{eq:app:gammatilde:kern}
  \tilde\gamma
  =
  \frac{s}{2} \,
  \frac{1}{\tho^2}
  \times
  \frac{2 \, (\hat{c}_0)^3}{\tho^2_{dw} \, (2-\nu \hat{c}_0)} \times \Im[\hat{I}_1 - \hat{c}_0 \, \hat{I}_2]
  \ ,
\end{equation}
and which indeed is rather unsimilar to the result obtained in Eq.~(\ref{eq:gammatilde:shearflow}) above.

We first recall the definitions (using the notation of \cite{Kern:2021}).
The variable $\hat{c}_0:=\frac{c_0}{\Uwind}$ is in fact simply the (monochromatic) wave age $\tho$, and the parameter $\nu$ was defined in \cite{Kern:2021} to characterise the vorticity as
\begin{equation}
  \label{eq:supp:nu}
  \nu := \Omega \, \Uwind/g
  \ .
\end{equation}
Also, the coefficients $\hat{I}_1$ and $\hat{I}_2$ are defined via the integrals
\begin{eqnarray}
  \hat{I}_1
  &:=& \frac{1}{\Uwind\, w_0} \int_{\zo}^\infty U_{ext}(z) \, w_a(z) \, k \, dz
  \\
  \hat{I}_2
  &:=& \frac{1}{w_0} \int_{\zo}^\infty  w_a(z) \, k \, dz
  \ ,
\end{eqnarray}
using the notation in \cite{Kern:2021}.

The correspondences are as follows:
$w=u_z$ in our notation, $w_0 = w(z{=}0)$ and $\tho_{dw} = \sqrt{g/k}/\Uwind$ is the (monochromatic) wave age for a deep water situation. The coefficients $\hat{I}_1$ and $\hat{I}_2$ can thus be expressed in the current notation as
\begin{eqnarray}
  \hat{I}_1  
  &=& \tho \int_{k\zo}^\infty \frac{U_{ext}(z)}{c} \, \frac{w_a(z)}{w_0} \, d(kz)
  \\
  \hat{I}_2  
  &=&  \int_{k\zo}^\infty 
  \, \frac{w_a(z)}{w_0} \, d(kz)
  \ ,
\end{eqnarray}
and therefore the sum of the expression in the brackets of Eq.~(\ref{eq:app:gammatilde:kern}) evaluates to 
\begin{equation}
\hat{I}_1 - \hat{c}_0 \, \hat{I}_2 = \tho \, \Io
\end{equation}
in our notation.

Substituting into the expression for $\tilde\gamma$ this yields, so far:
\begin{equation}
  \label{eq:gammatilde:sofar}
  \tilde\gamma = \frac{\tho^2}{\theta_{0,dw}^2 \, (2-\nu \tho)} \, s \, \Im[\Io]
  \ .
\end{equation}

We now address the parameter $\nu$ characterising vorticity, defined via Eq.~(\ref{eq:supp:nu}). 
Introducing the shorthand $\tilde{T}=\tanh(kh)$ we have, from Eq.~(63) in \cite{Kern:2021}, that
\begin{equation}
  \frac{\tho}{\sqrt{1-\nu \, \tho}}
  = \sqrt{\tilde{T}} \, \theta_{0,dw}
  \ ,
\end{equation}
which, upon substituting into Eq.~(\ref{eq:gammatilde:sofar}), leads to
\begin{equation}
  \label{eq:app:tmp}
  \tilde\gamma = \tilde{T} \, \frac{1-\nu\tho}{1-\frac{\nu\tho}{2}} \times \frac{1}{2} \Im[s \, \Io]
  \ .
\end{equation}

From this we can deduce an expression in terms of our vorticity parameter $\tilde{\Omega} = \frac{\Omega}{\omega_0}$ by transforming
\begin{equation}
  \nu \tho = \frac{\Omega \, c}{g}
  = \tilde{\Omega} \, \frac{k c^2}{g}
  = \tilde{\Omega} \, \left(\frac{c}{c_M}\right)^2
  = \tilde{T} \, \tilde{\Omega} \, (1-\nu\tho)
\end{equation}
so that
\begin{equation}
\nu\tho = \frac{\tilde\Omega \, \tilde{T}}{1+\tilde\Omega \, \tilde{T}}
\ ,
\end{equation}
and hence 
\begin{equation}
  \frac{1-\nu\tho}{1-\frac{\nu\tho}{2}}
  = \frac{1}{1+\frac{\tilde\Omega \, \tilde{T}}{2}}
  \ , 
\end{equation}
which ultimately transforms Eq.~(\ref{eq:app:tmp}) into
\begin{equation}
  \tilde\gamma = \frac{\tilde{T}}{1+\frac{\tilde{\Omega} \tilde{T}}{2}} \times \tilde\gamma_M
\ ,
\end{equation}
\ie we recover Eq.~(\ref{eq:gammatilde:shearflow}), as we have set out to show.

\end{document}